\documentclass[aps,pra,twocolumn,showpacs]{revtex4}

\usepackage{graphicx}
\usepackage{amsmath}

\begin{document}

\title{Momentum distribution and correlation function of
quasicondensates in elongated traps}
\author{F. Gerbier}
\email[e-mail: ]{fabrice.gerbier@iota.u-psud.fr}
\author{J. H. Thywissen}
\altaffiliation[current address:]{Department of Physics,
University of Toronto, Canada.}
\author{S. Richard}
\author{M. Hugbart}
\author{P. Bouyer}
\author{A. Aspect}
\affiliation{Laboratoire Charles Fabry de l'Institut
d'Optique\footnote{UMRA 8501 du CNRS}, 91403 Orsay, France}
\date{\today}

\begin{abstract}
We calculate the spatial correlation function and momentum
distribution of a phase-fluctuating, elongated three-dimensional
condensate, in a trap and in free expansion. We take the
inhomogeneous density profile into account via a local density
approximation. We find an almost Lorentzian momentum distribution,
in stark contrast with a Heisenberg-limited Thomas-Fermi
condensate.
\end{abstract}

\maketitle

%
%
Low-dimensional, degenerate Bose gases are expected to have
significantly different coherence properties than their
three-dimensional (3D) counterparts. In one-dimensional (1D)
uniform systems, no true condensate can exist at any temperature T
because of a large population of low-lying states that destroys
phase coherence (see \cite{petrov1d} and references therein). For
a trapped gas, the situation is different: the finite size of the
sample naturally introduces a low-momentum cutoff, and at
sufficiently low temperature $T \ll T_\mathrm{\phi}$, a phase
coherent sample can exist \cite{petrov1d}. Above
$T_\mathrm{\phi}$, the degenerate cloud is a so-called
{\it{quasicondensate}}: the density has the same smooth profile as
a true condensate, but the phase fluctuates in space and time. As
shown in \cite{petrov3d}, this analysis holds also for 3D
condensates in elongated traps even if, strictly speaking, radial
motion is not frozen. Such 3D, phase-fluctuating condensates have
been recently observed experimentally in equilibrium
\cite{dettmer01} and nonequilibrium \cite{schvarchuck02} samples.

Phase fluctuations of the condensate are caused mainly by
long-wavelength (or low-energy) collective excitations
\cite{petrov1d,petrov3d,landaups2}. In elongated traps, the lowest
energy modes are 1D excitations along the long axis of the trap
\cite{stringari98}. Furthermore, in the long-wavelength limit,
density fluctuations are small and can be neglected for the
calculation of the correlation function \cite{petrov1d,refstoof}.
Then, the single-particle density matrix is, assuming cylindrical
symmetry,
\begin{equation} \label{eq1}
\langle \hat{\Psi}^{\dagger}(\rho,z) \hat{\Psi}(\rho,z')\rangle
\approx  \chi(\rho,Z,s) e^{-\frac{1}{2} \Delta \phi^{2}(Z,s)} \; .
\end{equation}
We have introduced $\Delta \phi^{2}(Z,s) = \langle[ \phi(z)-
\phi(z')]^{2}\rangle$, the variance of the phase difference
between two points $z$,$z'$ on the axis of the trap, with mean
coordinate $Z=(z+z')/2$ and relative distance $s=z-z'$, and the
overlap function $\chi=\sqrt{n_0(\rho,z) n_0(\rho,z')}$, where
$n_0$ is the (quasi)condensate density. The variance $\Delta
\phi^{2}(Z,s)$, the key quantity to characterize the spatial
fluctuations of the phase of the condensate, has been calculated
in \cite{petrov3d}, and an analytical form has been given, which
is valid near the center of the trap ({\it{i.e.}} for $Z,s\ll L$,
with $L$ the condensate half-length). The first goal of this paper
is to find an analytical approximation for the variance $\Delta
\phi^{2}(Z,s)$ valid across the whole sample. This is motivated by
the fact that coherence measurements with quasi-condensates
\cite{richard03,hellweg03} are quite sensitive to the
inhomogeneity of the sample. In position space, interferometry
\cite{hagley99,bloch00} gives access to the spatial correlation
function $\mathcal{C}^{(1)}(s)$ (see {\it{e.g.}}
\cite{cctcargese})
\begin{equation} \label{eq2}
\mathcal{C}^{(1)}(s)= \int d^{3}{\bf{R}} \langle
\hat{\Psi}^{\dagger} (\rho,Z+s/2) \hat{\Psi}(\rho,Z-s/2) \rangle.
\end{equation}
Equivalently, one can measure the axial ({\it{i.e.}} integrated
over transverse momenta) momentum distribution $\mathcal{P}
(p_\mathrm{z})$, which is the Fourier transform of
$\mathcal{C}^{(1)}(s)$
\cite{stenger99,cctcargese,zambelli00,luxat03}:
\begin{equation} \label{eq3}
\mathcal{P}(p_\mathrm{z})=\frac{1}{2 \pi \hbar} \int ds \;
\mathcal{C}^{(1)}(s) e^{-i p_\mathrm{z} s / \hbar} \ .
\end{equation}
A powerful tool to measure $\mathcal{P}(p_\mathrm{z})$ is Bragg
spectroscopy with large momentum transfer, as demonstrated in
\cite{stenger99} for a 3D condensate, and recently applied in our
group to perfom the momentum spectroscopy of a quasicondensate
\cite{richard03}. It is clear that both $\mathcal{C}^{(1)}$ and
$\mathcal{P}$ are sensitive to the inhomogeneity of the system.
Our second goal is to obtain explicit expressions for these two
important quantities.

This paper is organized as follows. First, we summarize the
results of \cite{petrov3d}, and give an energetic interpretation
of $T_\mathrm{\phi}$. Next, we discuss in detail a local density
approach (LDA) to compute the variance of the phase for any mean
position in the trap. This approximation is found to be accurate
for $T \geq 8 T_\mathrm{\phi}$, when applied to a trapped
condensate. Using the LDA, we then address the problem of a
phase-fluctuating condensate in free expansion. In particular, we
point out that at higher temperature, the phase fluctuations
dominate over the mean-field release velocity and govern the shape
of the momentum distribution.

We consider $N_0$ condensed atoms, trapped in a cylindrically
symmetric, harmonic trap with an aspect ratio
$\lambda=\omega_{\mathrm{z}}/\omega_{\perp} \ll 1$. If $\mu \gg
\{\hbar \omega_{\perp}$,$\hbar \omega_{\mathrm{z}}\}$, the
condensate is in the 3D Thomas-Fermi (TF) regime \cite{dalfovo99}.
The density has the well-known inverted parabola form: $n_0
({\bf{r}}) = n_{\mathrm{0m}} (1-\tilde{\rho}^2 - \tilde{z}^2)$,
with the peak density $n_{\mathrm{0m}}=\mu /g$ related to the
chemical potential $\mu$. From now on, we use the reduced
coordinates $\tilde{\rho}=\rho/R$ and $\tilde{z}=z/L$, with
$R^{2}=2 \mu /M \omega_{\perp}^{2}$ and $L^{2}=2 \mu /M
\omega_{\mathrm{z}}^{2}$ respectively.

As shown in \cite{petrov1d}, phase fluctuations in trapped gases
are mostly associated with thermally excited, low-energy
quasi-particles (the quantum fluctuations are negligible). Under
these conditions, the variance $\Delta
\phi^{^2}(\tilde{Z},\tilde{s}) $ is
\begin{equation} \label{eq4}
 \Delta \phi^{^2}(\tilde{Z},\tilde{s})  \approx
\sum_{j} \frac{2 k_{\mathrm{B}}T}{\hbar \omega_j} |
\phi_{j}(\tilde{Z}+\tilde{s}/2)- \phi_{j}(\tilde{Z}-\tilde{s}/2)
|^{2},
\end{equation}
where the sum extends over the 1D axial excitations, with energy
$\hbar \omega_j$ and occupation number $N_j \approx
k_{\mathrm{B}}T/\hbar \omega_j$ for $N_j \gg 1$. For a 3D
condensate in an elongated trap, the amplitude $ \phi_{j}$ is
proportional to a Jacobi polynomial $P_{j}^{(1,1)}$, and
$\omega_{j}=\omega_{\rm{z}}\sqrt{j(j+3)}/2$ \cite{stringari98} for
integer $j$. The explicit result for the variance is then
\cite{petrov3d}
\begin{equation} \label{eq5}
 \Delta \phi^{^2}(\tilde{Z},\tilde{s})  =
\frac{T}{T_{\mathrm{\phi}}} f(\tilde{Z},\tilde{s}),
\end{equation}
with $f(\tilde{Z},\tilde{s})= \sum_{j} F_{j}
 [P_{j}^{(1,1)}
(\tilde{Z}+\tilde{s}/2)-P_{j}^{(1,1)}
(\tilde{Z}-\tilde{s}/2)]^{2}$, and the coefficients
$F_{j}=(j+2)(j+3/2)/4j(j+1)(j+3)$. Below the characteristic
temperature $T_\mathrm{\phi}$ $=$ $15 N_0 (\hbar
\omega_{\mathrm{z}})^{2}/32\mu k_{\rm{B}}$, the phase profile is
almost flat, and the single-particle density matrix (\ref{eq1}) is
limited by the overlap function $\chi$: therefore the
characteristic width of $\mathcal{C}^{(1)}$ ({\it{i.e.}} the
coherence length) is of order $L$. On the other hand, if $T \gg
T_\mathrm{\phi}$ the variance $\Delta \phi^{^2}$ dominates the
behavior of $\mathcal{C}^{(1)}$, and the coherence length is
substantially smaller than $L$. Near the center of the trap
($\tilde{Z},\tilde{s}\ll1$), Petrov {\it{et al.}} \cite{petrov3d}
have derived the simple law $ \Delta
\phi^{^2}(\tilde{Z},\tilde{s}) \approx (T/T_{\phi})|\tilde{s}|$,
and introduced the characteristic phase coherence length
$L_\mathrm{\phi}= L T_\mathrm{\phi}/T$ that depends implicitly on
the temperature, on the number of condensed atoms and on the
trapping geometry.

We can understand this expression for $L_\mathrm{\phi}$ from
energetic considerations. A random phase gradient of the
condensate wavefunction, on a length scale $L_{\phi}$, requires an
average kinetic energy $E_\mathrm{\phi} \sim N_0 \hbar^{2}/M
L_\mathrm{\phi}^{2}$. This kinetic energy is supplied by the
thermal excitations that drive the fluctuations of the phase
\cite{landaups2}. As these excitations are quasi-classical
($N_{k}\gg 1$), this energy is of order $ k_{\mathrm{B}}T$ times
the number of relevant modes. In 1D $k$-space, the distribution of
the relevant excitations extends over $ \sim 1/ L_\mathrm{\phi}$,
and the spacing between modes is $\sim 1/L$ because of the finite
size of the system: this gives $L/L_\mathrm{\phi}$ relevant modes.
By equating the two expressions for $E_{\phi}$, we recover finally
$ L_\mathrm{\phi}\sim L N_0(\hbar \omega_{\mathrm{z}})^{2}/\mu
k_{B} T$.

As indicated earlier, it is important to take the full spatial
dependence of $\Delta \phi^{^2}(\tilde{Z},\tilde{s}) $ into
account for quantitative comparison with experiments. In any case,
Eq.~(\ref{eq5}) can be evaluated numerically. However, we gain
physical insight with an analytical approach based on the local
density approximation (LDA), also used in \cite{dettmer01} to
calculate the evolution of the density in time-of-flight. This
approximation considers that the condensate is locally equivalent
to a homogeneous medium, however with a slowly varying density
that depends on the trapping potential. If $T \gg
T_\mathrm{\phi}$, the coherence length is sufficiently small
compared to $L$, that the LDA is valid for the calculation of
correlation properties.

The first step is to consider a finite cylinder of length $2L$,
with radial harmonic trapping and periodic boundary conditions
along $z$ (and therefore homogeneous axial density). For this
geometry, we find in the TF regime:
$n_0({\bf{r}})=n_{\mathrm{0m}}(1-\tilde{\rho}^{2})$ for the
condensate wavefunction. Low-lying excitations are found using
standard Bogoliubov theory \cite{landaups2}, after averaging over
the transverse degrees of freedom \cite{stringari98}. The
Bogoliubov spectrum for the excitation frequencies is
$\omega_k^{\mathrm{B}}=(\omega_k(\omega_k+2 M
c_{\rm{1d}}^{2}/\hbar))^{1/2} \approx c_{\mathrm{1D}}k$ for small
$k$, with the free particle energy $ \hbar \omega_k=\hbar^{2}
k^{2}/2M$ and the 1D speed of sound
$c_{\mathrm{1D}}=\sqrt{\mu/2M}$ \cite{zaremba98}. The Fourier
component for phase fluctuations with wavevector $k$ is
\begin{equation} \label{eq6} \phi_k =  \sqrt{\frac{
\omega_k^{\mathrm{B}}}{2\omega_k }} \frac{1}{\sqrt{\mathcal{V}}}
\approx \sqrt{\frac{Mc_{\mathrm{1D}}}{\hbar k}}
\frac{1}{\sqrt{\mathcal{V}}},
\end{equation}
where the final expression holds for low-lying phonon states ($k
\rightarrow 0$), and $\mathcal{V}=2 \pi n_{\mathrm{0m}} R^2 L$. In
a second step, we take into account the trapping potential by the
substitution:
\begin{equation} \label{eq7}
\mu \rightarrow \mu - \frac{1}{2}M \omega_{\mathrm{z}}^{2}z^{2}.
\end{equation}

This implies directly the replacements:
\begin{equation}  \label{eq8}
\begin{array}{lrcl}
{\mathrm{density}}: & n_{\mathrm{0m}} (\propto \mu) & \rightarrow & n_{\mathrm{0m}}(1-\tilde{z}^{2})\\
{\mathrm{speed\ of\ sound}}: & c_{\mathrm{1D}} (\propto \sqrt{\mu}) & \rightarrow & c_{\mathrm{1D}}\sqrt{1-\tilde{z}^{2}}\\
{\mathrm{radius}}: & R (\propto \sqrt{\mu}) & \rightarrow & R \sqrt{1-\tilde{z}^{2}}\\
{\mathrm{half-length}}: & L  & \rightarrow & L\\
\end{array}
\end{equation}
With these substitutions, we recover the 3D TF density profile. We
require that the excitation frequency $c_{\mathrm{1D}} k $ is not
modified as well, which implies replacing $k$ with
$k(1-\tilde{z}^{2})^{-1/2}$, and using a density  of states
$\mathcal{N}(k) dk = (L/ \pi) (1-\tilde{z}^{2})^{-1/2} dk $. For
the position-dependent variance of the phase, we find
\cite{jacobi}
\begin{equation} \label{eq9}
 \Delta \phi^{^2}(\tilde{Z},\tilde{s})  \approx
\frac{T}{T_{\mathrm{\phi}}} \frac{\mid \tilde{s}
\mid}{(1-\tilde{Z}^2)^{2}} \ .
\end{equation}
The $\tilde{Z}$-dependent phase coherence length $L_\mathrm{\phi}
(1-\tilde{Z}^2)^{2}$ appearing in Eq.~(\ref{eq9}) can be
substantially smaller near the edges of the trap than in the
center. We will see that this reduces the average coherence length
below its value at the center of the trap.

\begin{figure}[ht!]

\includegraphics[width=8cm]{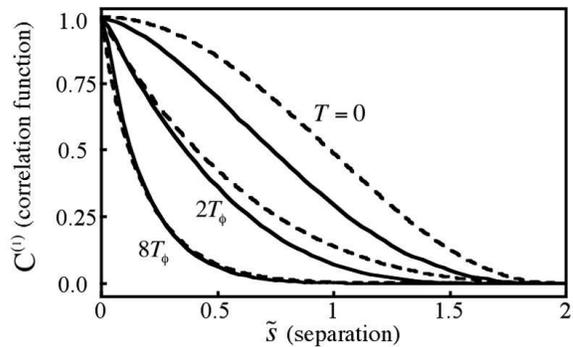}
\caption{Spatial correlation function
$\mathcal{C}^{(1)}(\tilde{s})$ of a trapped quasicondensate. The
solid lines follow from the numerical evaluation of the result
(\ref{eq5}) derived in \cite{petrov3d}, for $T= 0$, $T/T_{\phi}=4$
and $T/T_{\phi}=8$, in order of decreasing width. The dashed lines
follow from the local density approximation Eq.(\ref{eq10}). Note
that for convenience, $\mathcal{C}^{(1)}(\tilde{s})$ as shown here
is normalized to $1$ rather than to $N_{0}$, differently from what
is done in the text. } \label{fig01}
\end{figure}

We deduce from Eqs.(\ref{eq1},\ref{eq2},\ref{eq9}) the correlation
function:
\begin{multline} \label{eq10}
\mathcal{C}^{(1)}_{\mathrm{trap,T}}(\tilde{s}) \approx  \frac{15
N_{0}}{8}\int_{0}^{\sqrt{1-\tilde{s}^2/4}} d
\tilde{z}(1-\tilde{z}^{2}-\tilde{s}^2/4)^{2}\\
\exp{(-\frac{T}{2T_\mathrm{\phi}}\frac{\mid \tilde{s}
\mid}{(1-\tilde{z}^{2})^{2}})}.
\end{multline}
In deriving Eq.~(\ref{eq10}), we have used the approximation for
the overlap function $\chi(\rho,Z,s) \approx (1-
\tilde{\rho}^{2}-\tilde{z}^2-\tilde{s}^2/4)$, valid near the
center of the trap. In Fig.~\ref{fig01}, we compare the result
(\ref{eq10}) to the correlation function following from the
numerical integration of (\ref{eq5}). In the $T=0$ limit, the
correlation function is limited by the overlap $\chi$. Because of
the approximate form of $\chi$, our result
$\mathcal{C}^{(1)}_{\mathrm{trap,T=0 }} \approx
N_{0}(1-(\tilde{s}/2)^2)^{5/2}$ is about $25\%$ too broad, and one
should rather use the gaussian approximation to
$\mathcal{C}^{(1)}$ derived in \cite{zambelli00}. As $T$
increases, $\mathcal{C}^{(1)}$ turns to an exponential-like
function, and our approximation approaches the numerical
calculation. For $T> 8 T_\mathrm{\phi}$, the LDA result is very
close to the numerical one (maximum error $\approx 3\%$). For $T
\gg T_\mathrm{\phi}$, Eq.~(\ref{eq10}) can be further simplified
by keeping the $\tilde{s}$-dependent term only in the exponential.
The Fourier transform then gives the momentum distribution
\begin{equation} \label{eq11}
\mathcal{P}_{\mathrm{trap,T}}(p_{\mathrm{z}})\approx \frac{15 N_{0}
p_\mathrm{\phi} }{32 \pi} \int_{-1}^{1} d \tilde{z}
\frac{(1-\tilde{z}^{2})^{4}}{(1-\tilde{z}^{2})^{4} p_{\mathrm{z}}^{2}
+ p_\mathrm{\phi}^{2}/4 },
\end{equation}
where $p_\mathrm{\phi}=\hbar/  L_\mathrm{\phi}$ is a typical
momentum associated with the phase fluctuations. This function is
self-similar in $p_{\rm{z}}/p_{\phi}$, and approximated to better
than 4\% by a normalized Lorentzian with a half-width at
half-maximum (HWHM) of $\Delta p = 0.67 p_\mathrm{\phi}$. This
Lorentzian shape of the momentum distribution differs
qualitatively from the fully coherent case, where it is almost
Gaussian and limited by the Heisenberg principle \cite{stenger99}.
The increase of the phase fluctuations with increasing $T$ not
only broadens the momentum distribution, but also induces the
appearance of "wings", that form the ``high-energy tail'' of the
quasicondensate. To quantify the accuracy of our approximation, we
have calculated numerically the Fourier transform of the
correlation function. We find empirically that the HWHM is
accounted for by the formula $\Delta p^{2} \approx
(2.04\hbar/L)^2+(0.65 \hbar/L_{\phi})^2$. The first term
corresponds to the Heisenberg-limited momentum width, and the
second to the phase fluctuations. For $T \geq 8 T_{\phi}$, the
height and width agree to better than 4\% with the Lorentzian
approximation. For lower $T$, the overlap function $\chi$ still
affects the momentum distribution.

Note finally that the momentum distribution is Lorentzian only in
the domain $k \ll R^{-1}$. Outside of this region, the 3D nature
of the excitations should to be taken into account properly.
However, this does not affect the quasicondensate peak we are
investigating here, but only the much smoother thermal background
\cite{stenger99,zambelli00}.

The results of the above paragraphs are valid for an equilibrium
situation. However, coherence measurements involving Bragg
scattering \cite{hagley99,stenger99} suffer from two major
difficulties in a very elongated trap \cite{richard03}: mean-field
broadening of the resonance \cite{stenger99}, and elastic
scattering from the recoiling atoms and the condensate towards
initially empty modes \cite{chikkatur00}. Both of these problems
can be solved by opening the trap abruptly, and letting the BEC
expand to decrease its density before measurement. In the
remainder of this paper, we discuss how expansion modifies the
momentum distribution and the correlation function, assuming that
the expansion time is chosen to be long enough to neglect the
collisions.

For a pure, elongated condensate abruptly released from the trap
at $t=0$, the explicit solution was found in \cite{kagan96}. The
condensate density keeps its initial Thomas-Fermi shape, with the
coordinates re-scaled. The (small) axial momentum from the
released mean-field energy is linear in position:
$p_{\mathrm{z}}\approx p_{\mathrm{exp}}\tilde{z}$, with
$p_{\mathrm{exp}}= (\pi/\sqrt{2}) \lambda M c_{\mathrm{S}}$ for
$\tau=\omega_{\perp} t \gg 1$, and $c_{\mathrm{S}}=\sqrt{\mu/M}$
is the 3D speed of sound. The axial momentum distribution mirrors
the (integrated) density distribution:
\begin{equation} \label{eq12}
\mathcal{P}_{\mathrm{exp,T=0 }}(p_{\mathrm{z}}) = \frac{15 }{16
p_{\mathrm{exp}}}
\left(1-(\frac{p_{\mathrm{z}}}{p_{\mathrm{exp}}})^2\right)^2.
\end{equation}
This expression holds for a pure condensate, at $T=0$, as
indicated. For a phase fluctuating condensate at finite T, it is
necessary to consider the time evolution of the fluctuations as
well. As shown in \cite{dettmer01}, the momentum distribution
partially converts into density modulations after time of flight.
An explicit solution was derived for the density fluctuations in
the axially homogeneous case. Using the continuity equation (after
radial averaging), we find for $\tau \gg 1$
\begin{equation} \label{eq13}
\phi_{\rm{k}}(z,\tau) \approx   \phi_k(z,0)
\tau^{-(\omega_k^{\mathrm{B}}/\omega_{\perp})^{2}}
\cos{(\frac{\omega_{\rm{k}}}{ \omega_{\perp}}\tau)} ,
\end{equation}
If $\omega_{\rm{z}} t \ll(\mu/\hbar
\omega_{\rm{z}})(T/T_{\phi})^{2}$, then for all $k \lesssim
L_\mathrm{\phi}^{-1}$, the phase distribution is essentially
frozen: $ \phi_{\rm{k}}(z,\tau) \approx \phi_{\rm{k}}(z,0)$.
Physically, this condition states that for such a time of flight,
the excitations that have significant contributions to the phase
fluctuations have not yet been converted into density modulations.
This condition is not at all restrictive for typical experimental
parameters \cite{dettmer01,schvarchuck02,richard03,hellweg03}, and
we suppose it is met in the remainder of the paper.

\begin{figure}[t]

\includegraphics[width=8cm]{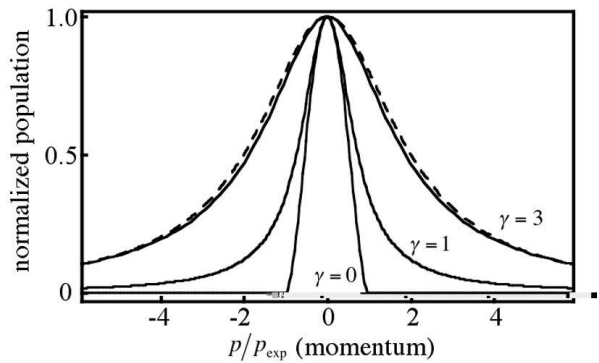}
\caption{The momentum distribution in expansion $g_{\gamma}$
(Eq.~\ref{eq16}) for
$\gamma=p_{\mathrm{\phi}}/p_{\mathrm{exp}}=0,1,3$. As $\gamma$
increases, $g_{\gamma}$ continuously transforms from a quartic
profile to a lorentzian-like profile (see text). For $\gamma=3$,
we find little change from the momentum distribution in the trap
(dashed line, a Lorentzian with HWHM $0.67 \gamma$, see text). The
functions have been rescaled by their maximum values to facilitate
comparison.} \label{fig02}
\end{figure}

Using the rescaled wavefunction from \cite{kagan96}, together with
(\ref{eq9}), we find the correlation function for the expanding
quasicondensate:
\begin{multline} \label{eq14}
\mathcal{C}^{(1)}_{\mathrm{exp,T}}(\tilde{s})\approx \frac{15
N_{0} }{16}\int_{-1}^{1} d \tilde{z}(1-\tilde{z}^{2})^{2}\\
\exp{(i \frac{\pi \mu}{\hbar
\omega_{\perp}}\tilde{z}\tilde{s}-\frac{T}{2T_\mathrm{\phi}}\frac{\mid
\tilde{s} \mid}{(1-\tilde{z}^{2})^{2}})} .
\end{multline}
The phase factor in (\ref{eq14}) accounts for the local expansion
momentum introduced above (recall $\pi \mu / \hbar \omega_{\perp}
\gg 1$). The Fourier transform gives the momentum distribution
\begin{equation} \label{eq15}
\mathcal{P}_{\mathrm{exp,T}}(p_{\mathrm{z}})\approx
\frac{N_{0}}{p_{\mathrm{exp}}}g_{\gamma=p_\mathrm{\phi}/p_{\mathrm{exp}}}(\frac{p_{\mathrm{z}}}{p_{\mathrm{exp}}}).
\end{equation}
The function $g_{\gamma}$ is given by
\begin{equation} \label{eq16}
g_{\gamma}(x)= \frac{15  \gamma }{32 \pi} \int_{-1}^{1} d
\tilde{z} \frac{ (1-\tilde{z}^{2})^{4}} {(1-\tilde{z}^{2})^{4}
(x-\tilde{z})^{2} + \frac{\gamma^{2}}{4} },
\end{equation}
and the ratio $\gamma=p_\mathrm{\phi}/p_ {\mathrm{exp}}$ controls
which component of the momentum distribution dominates. In the
limit $T \rightarrow 0$, using $ \gamma /(x^{2}+\gamma^{2})
\rightarrow \pi \delta(x)$ as $\gamma \rightarrow 0 $, we recover
the zero-temperature result Eq.~(\ref{eq12}). On the other hand,
if $\gamma \gg 1$, we expect the momentum distribution to be
similar to the distribution in the trap (\ref{eq11}).
Fig.~\ref{fig02} shows a numerical calculation of $g_{\gamma}$,
for various values of $\gamma$. We find that already for $p_{\phi}
\gtrsim 2 p_{\rm{exp}}$, the momentum distribution is almost
entirely dominated by phase fluctuations, and, as in the trapped
case, is very well approximated by a normalized Lorentzian with
HWHM $= 0.67 p_{\phi}$. Here, we note two points: first, that the
Heisenberg width $\sim \hbar/L$ is negligible at any temperature,
and second that, for large enough condensates, we can have
$p_{\rm{exp}} \gg p_{\phi}$ even if the coherence length is
smaller than $L$.

In summary, we have analyzed the measurement of phase fluctuations
in elongated Bose condensates. Within a local density approach, we
have been able to take the density profile into account, and
derived analytical formulas for the correlation function and the
momentum distribution of static and freely expanding
quasicondensates. In the regime of interest, the formula compare
well to a numerical evaluation based on the results of
\cite{petrov3d}, which are exact in the long-wavelength limit. In
particular, we show how the shape of the momentum distribution
tends to a Lorentzian with half-width $\approx 0.67
\hbar/L_{\phi}$ as one goes further in the phase-fluctuating
regime. We believe that these results may be helpful to understand
quantitatively experiments involving quasi-condensates
\cite{dettmer01,schvarchuck02,richard03,hellweg03}.

We gratefully acknowledge D. S. Petrov and G. V. Shlyapnikov for
many stimulating discussions, for useful comments on the
manuscript and also for providing us with a detailed version of
the calculations published in \cite{dettmer01}. JHT is supported
by a Chateaubriand Fellowship and MH by IXSEA. This work is
supported by the EU and DGA.

\begin{thebibliography}{lll}
\bibitem{refstoof} It is necessary to improve Bogolubov
theory \cite{andersen02,mora02} to obtain the equation of state in
any dimensions, in presence of phase fluctuations. For the
specific problem we address in this paper, the long wavelength
formulation is sufficient.
\bibitem{jacobi} We have also verified that, starting from the large $j$ limit of the
Jacobi polynomials, one can recover Eq.~(\ref{eq9}).
\end{thebibliography}
\end{document}